\documentstyle[prb,aps,twocolumn,epsf]{revtex} 

\begin{document}
\title{Inhomogeneous magnetism in single crystalline Sr$_3$CuIrO$_{6+\delta}$: 
Implications to phase-separation concepts}
\author{Asad Niazi, P.L. Paulose and E.V. Sampathkumaran$^*$}

\address{Tata Institute of Fundamental Research, Homi Bhabha Road, Colaba,
Mumbai - 400005, India}

\maketitle

\begin{abstract}

The single crystalline form of an insulator,
Sr$_3$CuIrO$_{6+\delta}$, is shown to exhibit unexpectedly more than one 
magnetic transition (at 5 and 19 K) with spin-glass-like 
magnetic susceptibility behaviour. On the basis of this finding, viz.,   
inhomogeneous magnetism in a chemically homogeneous 
material, we propose that the idea of "phase-
separation" described for manganites\cite{1} is more widespread in different ways.
The observed experimental features enable us to make a comparison with the predictions of a recent 
toy model\cite{2} on {\it magnetic} phase separation in an insulating environment.  
\end{abstract}
\vskip0.5cm
$^*$Corresponding author. E-mail address: sampath@tifr.res.in
\vskip1cm
\maketitle

The concept of electronic phase separation has been recently advocated
to describe spin-glass-like characteristics induced by disorder in
doped manganites and such a state is even thought of as a new state of
matter.\cite{1}  An important question\cite{1} remaining unaddressed is whether the
anomalies characterizing
"phase-separation" phenomenon 
can be seen even in those
{\it homogeneous} materials (that is, without any additional chemical doping) 
where there is {\it no coexistence} of localised
and itinerant charge carriers characterizing these manganites. In fact, recently, a toy model was also 
proposed\cite{2} predicting the role of disorder on the behavior 
of a system in which any 
two phases in general, not necessarily of insulator and metal,  
compete and it is of interest to provide experimental data in different situations 
to test this
theory.     In this Letter, we report the magnetic behavior 
of a pseudo-one-dimensional {\it insulator},\cite{3,4,5,6}
Sr$_3$CuIrO$_6$, in the single crystalline form. 
The observed data  reveal that the magnetism is of an
inhomogeneous nature  in these single crystals, with a broad 
agreement of the experimental features with the predictions of the toy model mentioned 
above.  Hence, we emphasize
that the phenomenon of "phase separation" -- within the criterion\cite{1} of
inhomogeneous magnetism in a homogeneous material -- is not restricted to
manganites alone.        

The compounds of the type, (Sr, Ca)$_3$MXO$_6$ (M, X = a metallic ion,
magnetic or non-magnetic), crystallizing in the K$_4$CdCl$_6$-type
rhombohedral structure (space group R$\bar3$c), are of considerable
importance due to the presence of spin-chains (M-X)
separated by Sr/Ca ions (see Refs. 3--13 and references cited therein).  
Among these, in the Cu containing compounds, there is a 
lowering of the crystal symmetry to monoclinic (space group 
C2/c) due to Jahn-Teller (JT) effect at the Cu site, a point of importance to
the discussion in this article.  
The readers may see Ref. 7 for more details on crystallographic features.
 
The single crystals of 3 to 5 mm length and 0.5 to 1 mm diameter were
grown\cite{14} by the flux method employing basic alkali fluxes.\cite{15} The
crystals were found to undergo monoclinic modification and to grow along
the [101] monoclinic axis using single-crystal x-ray diffraction. The lattice
constants were found to be: a = 9.298(2) \AA, b= 9.714(2) \AA, c =
6.710(2)\AA\ and $\beta$ = 92.19(2)$^{\circ}$. For two orientations of the crystal,
viz., H//[101] and H$\perp$[101], we have measured dc magnetic susceptibility ($\chi$) as a function
of T (2--300 K) at different magnetic fields (H = 10, 100 and 5000 Oe) for
the zero-field-cooled (ZFC) and field-cooled (FC) states of the specimens
by commercial magnetometers, superconducting quantum interference device
(Quantum Design) as well as vibrating sample magnetometer (Oxford
Instruments). In addition, ac $\chi$ measurements were performed at low
temperatures (2--25 K) at various frequencies. 
 
The T-dependence of dc $\chi$ below 25 K is shown in Fig. 1 for the ZFC
and FC states of the crystal. On comparing the magnitudes of the
magnetization for the two different orientations mentioned above, we can
conclude that the easy axis of magnetization lies perpendicular to the [101]
direction. This is in agreement with the conclusions based on neutron
diffraction pattern of a similar Cu compound.\cite{16} The low-field (H = 10
and 100 Oe) data show that distinct magnetic ordering appears at two temperatures, 
5 K (T$_1$) and 19 K (T$_2$), the former being much more prominent. The intensity (at low
fields) at T$_2$ is nearly the same for both the directions, whereas at T$_1$
it is markedly enhanced for H$\perp$[101] as compared to H//[101]. This
suggests that there is a significant magnetic anisotropy only for the T$_1$
transition.  For comparison, in the same figure we also display the low-field
behavior of the polycrystalline form reported earlier.\cite{5} In contrast to
single crystal data, in polycrystals the 19 K transition is very prominent while
that at 5 K appears only as a shoulder in the low field ZFC data; this 
transition gets more pronounced for moderate values of H.\cite{3,5,6}  
The fact that this crystal  is characterized by more than one magnetic
transition is endorsed by similar features in the ac $\chi$ data (Fig. 2). 

We now address the question of the nature of the magnetic ordering at T$_1$
and T$_2$. The real part of ac $\chi$ data (Fig. 2) reveals a spin-glass-like frequency
dependence with a marginal shift of the curves to higher temperatures
 with increasing frequency. This is
observed for both the transitions, implying that the magnetic structure is
frustrated both at T$_1$ and T$_2$. In support of this, we note that
there is a significant difference between ZFC and FC $\chi$ vs T curves at
low fields for both the transitions (Fig. 2). We have also observed well-defined upturn 
in the imaginery part of ac $\chi$ at these transitions with a similar frequency dependence, 
a feature typical of only magnetically frustrated
systems, and due to space limitations we do not present these data. 

The presence of more than one magnetic transition with spin-glass-like
behavior is the crucial finding for our main conclusion. 
It may be noted that all the octahedral (as well as trigonal-prismatic)
sites are crystallographically equivalent in the
undistorted structure and that spin-glass behavior is not
necessarily a favored state as the analogous Zn compound Sr$_3$ZnIrO$_6$
(Ref. 6) is antiferromagnetic ({\it vide infra}). The reversal of relative
intensity of these two transitions for single crystals when compared with the
polycrystalline behavior also clearly excludes the possibility of a single
magnetic transition showing progressive spin reorientation effects with a 
change in temperature. The features attributable to this phenomenon should be 
intrinsic to the material and should not depend upon single/poly-crystalline physical 
states. The same argument can be given to rule out possible independent 
ordering of Cu and Ir at T$_1$ 
and T$_2$; this is also inferred from the observation of unequal 
magnitudes of $\chi$ at these transitions even though spin 
of each of  these ions is  = 1/2. However, our conclusion is not 
dependent on whether both these ions order simultaneously or separately (as 
a given magnetic segment proposed below can contain either of these or both).  
In view of this situation, we propose that this single crystal could be classified as a
{\it magnetically phase separated system}\cite{1} in the sense that {\it there is a
coexistence of magnetic-segments, isolated by
defects, ordering magnetically at different temperatures.} 
If these segments are strongly magnetically
coupled in a homogeneous manner, one should have expected only one
magnetic transition of a long-range type in contrast to the observations.

This scenario prompts us to view the present finding in light 
of the predictions of a recent toy
model,\cite{2} relevant points of which are briefly outlined now. 
In order to extend the concept of phase separation to situations more
general than manganites, Burgy et al\cite{2} explored the consequences of 
a competition between two ordered states (separated by first order
transition) in the presence of disorder within 
a toy model; this model does not even assume 
itinerancy or localisation of the carriers and in this sense it is more general.
This theory predicts  that, below a characteristic temperature, 
there is an intermediate temperature range with
preformed clusters, but with uncorrelated order parameters, giving a globally 
paramagnetic state; in the lower-T regime, the clusters grow in size although
the disorder is uncorrelated and percolate upon cooling eventually resulting in long
range ordering; an application of a small magnetic field should have a dramatic
influence on initial isothermal magnetization (M) followed by non-saturation
at higher fields in the event of a competition between
ferromagnetic and antiferromagnetic phases.

In light of above model, we now look at the magnetic behavior of this material, both in 
polycrystals\cite{5,6} as well as in single crystal (see Fig. 3, top). 
The plots of inverse $\chi$ versus T are found
to be non-linear. One can see this dramatically in single 
crystals for H//[101] (see Fig. 3, top)  in the sense that there is a steep, 
but gradual, fall of inverse $\chi$ below 100 K, though the non-linearity
persists even for the perpendicular direction. This implies varying size of 
the magnetic cluster with decreasing temperature.
This increasing intercluster coupling is apparently manifested as a 
weak magnetic transition at about 19 K in the single crystal, 
with the rest being paramagnetic at this temperature. 
The intensity of the response at 19 K however must depend upon the number of such clusters. 
As the disorder in the polycrystalline form is expected to be more, one expects 
large number of such independent clusters, and hence the response for the polycrystalline 
form\cite{5} is
relatively larger compared to the single crystalline form (see Fig. 1). With a further
decrease of T, true bulk ordering develops around 5 K. 
There is also a steep increase of M (from zero in
zero H) by a small
application of magnetic field at 2 K without any saturation at higher fields.  
These observations are in 
accordance  with the above theory. 
However, there are some difficulties in claiming that there is a total agreement  
with respect to this model.  For instance, (i) we are 
unable to precisely pinpoint the characteristic temperature, whether it is 100 K or 19 K,
as the magnetic cluster formation appears to be a continuous process with decreasing
temperature and (ii) the magnetic ordering at T$_1$ appears to be more of a spin-glass-like, 
rather than of a long range type. However, 
we believe such details may be redundent for the present purpose; real systems can behave in a 
slightly different fashion depending on other complexities 
(e.g., the degree of disorder, the way the segments grow and couple with a variation of T, 
crystallographic arrangement, dimensionality etc) of the
systems. In our case it appears that the competition regime occurs in a more 
extended T range than assumed by the theory. 
What is important is that this system catches essential physics of
the model in terms of how the magnetic interaction among the competing 
"magnetic phases" evolve with decreasing
temperature in the presence of disorder in a homogeneous material. 
Finally, we like to stress that this compound is in fact 
on the verge of long range magnetic order as elaborated in Ref. 5, since
small applications of H (as low as 1 kOe) depress spin-glass characteristics. This is found
to be true for single crystal as well. Therefore, the discrepancy (ii)
is not a serious one and it is of interest to focus future studies to subject this system to other
small perturbations (by chemical or external pressure?) to stablise this long range
 magnetic phase, so as to demonstrate 
the behavior of this system closer to the predictions of the theory.

We now briefly 
discuss how defects (creating segments) can occur in this class of compounds and the
reader can see Ref. 17 for details. 
Apart from vacancies along the magnetic chains, the defects can also
arise due to the movement of a small fraction of Sr
ions to vacant sites in the chain  and it has been shown that the degree of
such defects could be tuned by varying synthetic conditions.
It has also been established that certain heat treatment conditions
increase the oxygen content\cite{5} 
and the extra oxygen (0$<$ $\delta$ $<$ 1) occupies a trigonal prismatic
site, resulting in the conversion of one such site into two octahedral sites. 
These kinds of defects are found to occur not only for Cu compounds, but also for 
Zn/Ni compounds and such "defective" forms 
have been called "incommensurate" phases.
In our earlier polycrystalline studies,\cite{5} for the Cu specimens with extra oxygen, 
the magnetic transition temperature has been found to be lower. For such reasons, 
we have denoted the oxygen content as "6+$\delta$" at least at few places
in this article.

Further comparison of the magnetic behavior of this Cu system with the
analogous Zn compound, Sr$_3$ZnIrO$_6$ (Ref. 6), is quite intriguing. 
As mentioned earlier,  the Zn compound undergoes long-range magnetic
ordering (antiferromagnetic, from Ir ions) at 19 K and the N\'eel temperature is
quite robust to synthetic conditions,\cite{5,6} unlike in the Cu compound, in spite
of the presence of defects.  
Therefore, an
additional crucial effect must be necessary in the Cu compound to make
the magnetism inhomogeneous. As discussed in Ref. 3, 
the JT active Cu ion results in multiple Cu-Cu and Cu-Ir
distances, which we believe is a crucial factor for
 randomizing magnetic interaction. However, 
 one can also argue that this distortion {\it alone} may cause a uniform, rather than
non-uniform, modification 
of the nearest neighbour Cu-Cu and Cu-Ir magnetic interaction 
throughout the crystal. Therefore, we propose that it is the presence of 
the defects that somehow 
separates out islands of magnetic ions 
("phase separation") of a given exchange interaction strength following JT distortion.  
We welcome future experiments to verify this 
conjecture.  

Various other interesting observations arise from our measurements. (i) The
peak in the low-field ZFC dc $\chi$ data at about 3 K vanishes at a higher field
(5 kOe), as does the ZFC-FC bifurcation. This implies complex 
thermomagnetic history effects on magnetism even in single crystals. (ii) 
The saturation moment (at 2 K) obtained by linear extrapolation 
of the high field data (60--120 kOe)
to zero field even for H parallel to easy axis 
($\sim 0.4 \mu_B$ for H$\perp$[101]) is much lower than that
expected for ferromagnetically-coupled Ir and Cu (S = 1/2) ions (see Fig. 3, bottom). 
The reduced value implies the existence of a moment-compensation effect, e.g., 
the magnetic frustration among the segments/chains. This behavior is the
same as in polycrystals.\cite{5} Further
evidences for frustration were presented in our earlier,\cite{5} e.g., 
by the absence of a heat-capacity
anomaly at the transition.  (iii) The ac $\chi$
data for H//[101] reveal that there are at least two transitions below 5 K
separated by about 1 K, with the vertical arrow in Fig. 3 marking one
transition and the peak representing another transition. A careful inspection of
the ZFC dc $\chi$ data at low fields also reveals the existence of a shoulder
at a temperature slightly above the peak. These observations suggest that
one gets additional transitions at different temperatures depending upon the
size of the segments regulated by defects/disorder as pointed out by Dagotto
et al.\cite{1} We find that the relative contribution to $\chi$ for the two
closely-spaced transitions near T$_1$ is also a sensitive function of H, as
revealed by reversal of the peak positions in dc and ac $\chi$ plots. In
addition, the transitions at T$_1$ and T$_2$ broaden out upon application of
a higher field (5 kOe) (Fig. 1), indicating strong field-induced intersegment
coupling effects.   (iv) For
H//[101], for the ZFC data at low fields, the value of $\chi$ for T$_1$ $<$ T
$<$ T$_2$ is negligibly small, indicating near-compensation of the magnetic
moments among the segments. This behavior too is modified by changing the
history of the sample (FC, higher fields, etc).  
All these observations, revealing the sensitivity of magnetically
separated segments to the magnetic field, history effects, etc., 
may be useful for further refinement of theoretical formulations.

To conclude, we have observed distinct evidence for inhomogeneous
magnetism in single crystalline
Sr$_3$CuIrO$_{6+\delta}$, on the basis of which the idea of phase-separation for 
a chemically homogeneous insulating medium is proposed.     Presumably, an interplay between
Jahn-Teller-effect and defects is required to result in inhomogeneous magnetism in this material. 
We have argued that this material
apparently serves as a favorable testing ground for a recent toy model of 
phase-separation\cite{2} 
in an insulating environment. It will be very fascinating if 
there is a consensus on the kind of phase-separation proposed in this article, 
as unlike in manganites (i) there is no coexistence of localised
and itinerant magnetic species and (ii) the "magnetic phases" are produced without
involving any chemical doping. In this sense, the root-cause of spin-glass-like anomalies in
this system is different from that in other conventional spin-glasses in which
chemical substitution is required for randomizing exchange coupling. In view
of this, we offer an explanation to a question raised\cite{1} in this regard: In which way 
are phase-separated materials different from spin-glasses? In our opinion, 
it is the "cause" which is responsible for the "result" that makes 
the former fundamentally different from the latter. Finally, we hope that this article
triggers more work on spin-chain oxides with this crystal structure. 

We thank Satish Ogale for his useful comments.  


\begin{figure}
\centerline{\epsfxsize=5cm{\epsffile{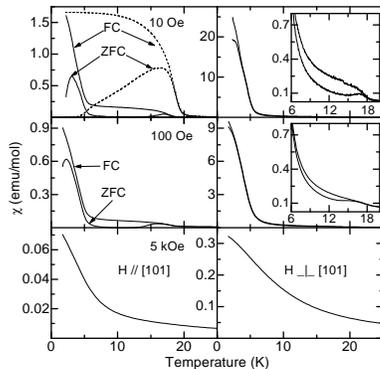}}}
\caption{Temperature dependence of dc magnetic susceptibility below 25 K
(H = 10, 100 and 5000 Oe) for the ZFC and FC states of single crystalline
Sr$_3$CuIrO$_6$ for two orientations. The ZFC-FC curves for H = 5000 Oe
overlap. The data (dashed line) at 10 Oe for the polycrystal from Ref. 5 are
included for comparison; note that the values are scaled down by a factor of
20 to accommodate in the same figure.  The insets highlight the 19 K
transition.}
\end{figure}
\begin{figure}
\centerline{\epsfxsize=5cm{\epsffile{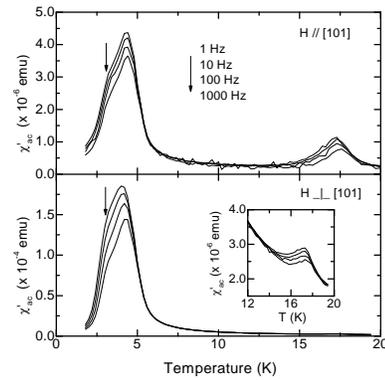}}}
\caption{Real part of ac susceptibility as a function of temperature for two
orientations of the Sr$_3$CuIrO$_6$ crystal at different frequencies in the
region of magnetic ordering. The vertical arrow marks an additional magnetic
transition below the peak. The inset (bottom figure) highlights the 19 K
transition.}
\end{figure}
\begin{figure}
\centerline{\epsfxsize=5cm{\epsffile{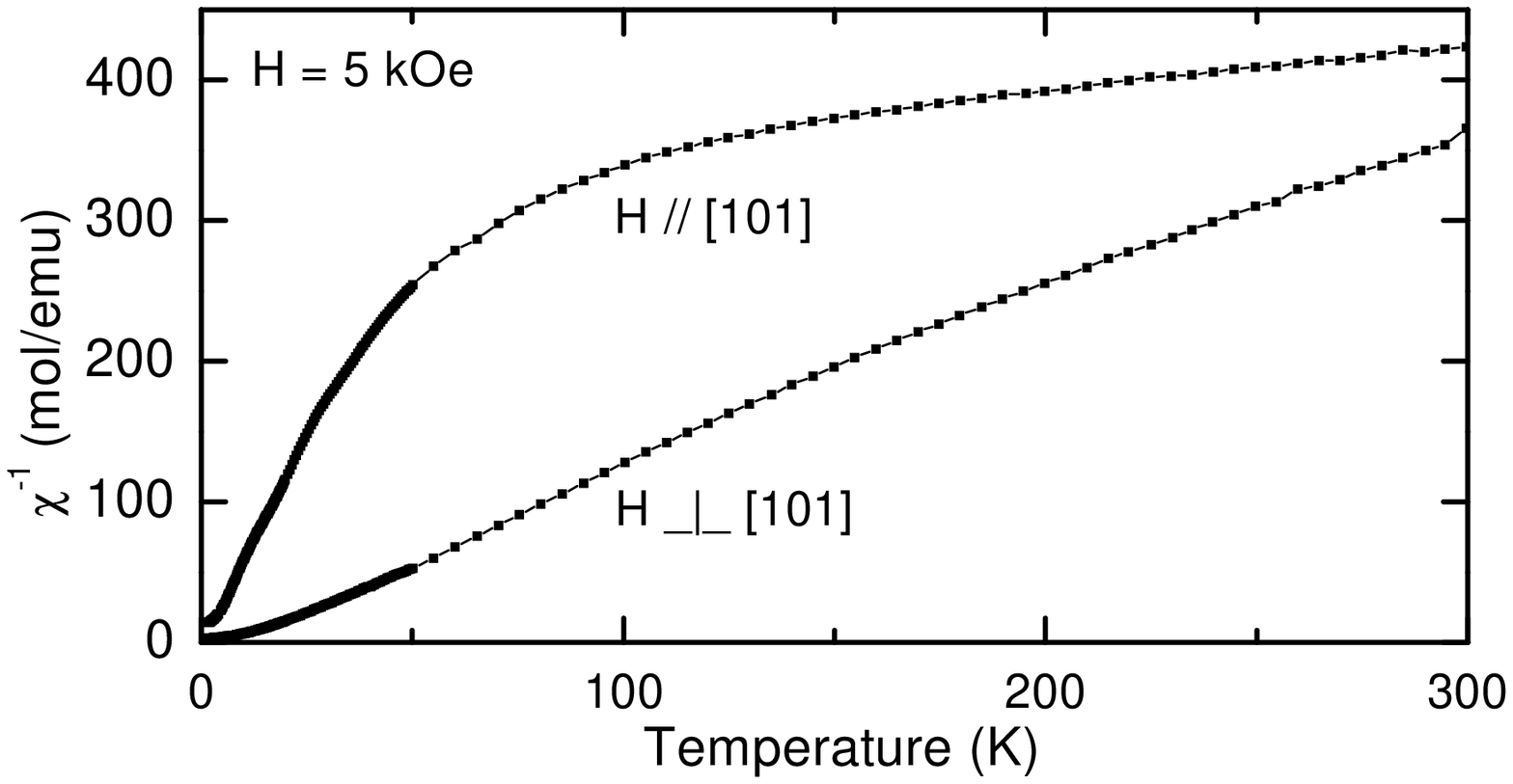}}}
\centerline{\epsfxsize=5cm{\epsffile{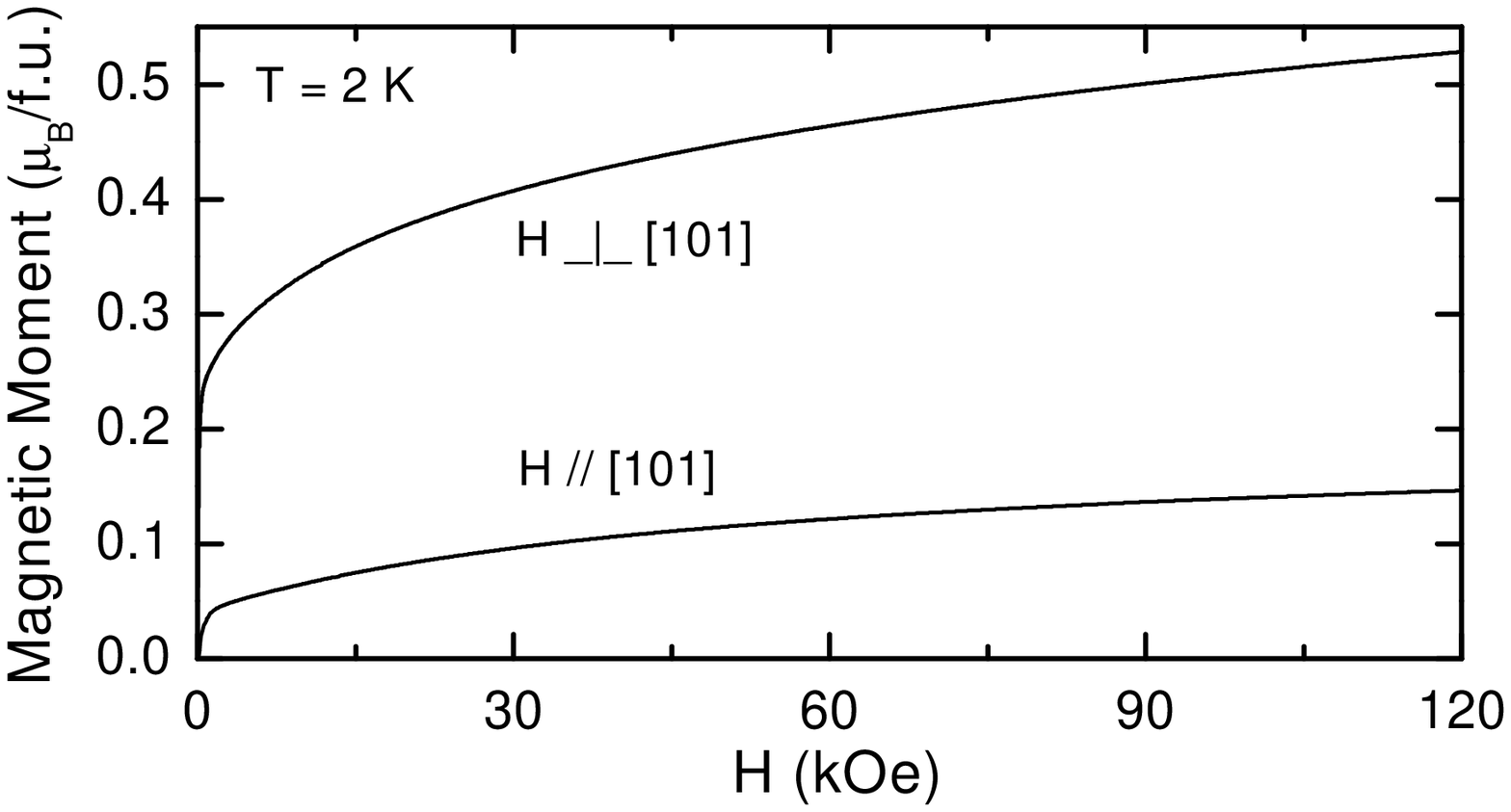}}}
\caption{Inverse susceptibility (H = 5 kOe, ZFC) (top) as a function of temperature
and isothermal magnetization at 2 K (bottom) 
for two orientations of the Sr$_3$CuIrO$_6$ crystal.}
\end{figure}
\end{document}